%% file: main.tex
\pdfoutput=1

\documentclass[12pt,a4paper]{article}

\usepackage{ifthen} 
\newboolean{pdflatex}
\setboolean{pdflatex}{true} 
\newboolean{articletitles}
\setboolean{articletitles}{true} 

\newboolean{uprightparticles}
\setboolean{uprightparticles}{false} 

\newboolean{inbibliography}
\setboolean{inbibliography}{false} 

\def\paperauthors{LHCb collaboration} 
\def\paperasciititle{Precision measurement of the Xicc++ mass} 
\def\papertitle{Precision measurement of the $\Xiccpp$ mass} 
\def\paperkeywords{{High Energy Physics}, {LHCb}} 
\def\papercopyright{\the\year\ CERN for the benefit of the LHCb collaboration} 
\def\paperlicence{CC-BY-4.0 licence}
\def\paperlicenceurl{https://creativecommons.org/licenses/by/4.0/}

\input{preamble}
\usepackage{longtable} 

\usepackage{float}%
\usepackage{graphicx}%
\usepackage{subcaption}%
\usepackage{amsmath}
\usepackage{wasysym}
\begin{document}

\renewcommand{\thefootnote}{\fnsymbol{footnote}}
\setcounter{footnote}{1}

\input{mass-result-macros.tex}

\input{title-LHCb-PAPER}


\renewcommand{\thefootnote}{\arabic{footnote}}
\setcounter{footnote}{0}



\pagestyle{plain} 
\setcounter{page}{1}
\pagenumbering{arabic}


\input{mass-paper-body.tex}

%


\addcontentsline{toc}{section}{References}
\bibliographystyle{LHCb}
\bibliography{main,standard,LHCb-PAPER,LHCb-CONF,LHCb-DP,LHCb-TDR,Xicc}


\clearpage
\input{LHCb_Authorship_09-Oct-2019.tex}

\end{document}

%% file: preamble.tex
\usepackage[top=1in, bottom=1.25in, left=1in, right=1in]{geometry}

\usepackage{microtype}
\usepackage{lineno}  
\usepackage{xspace} 
\usepackage{caption} 

\usepackage{graphicx}  
\usepackage{color}
\usepackage{colortbl}
\graphicspath{{./figs/}} 
\DeclareGraphicsExtensions{.pdf,.PDF,png,.PNG}

\usepackage{amsmath} 
\usepackage{amssymb}
\usepackage{amsfonts}
\usepackage{upgreek} 
\usepackage{listings}
\usepackage{paralist}

\newcommand*\patchAmsMathEnvironmentForLineno[1]{%
\expandafter\let\csname old#1\expandafter\endcsname\csname #1\endcsname
\expandafter\let\csname oldend#1\expandafter\endcsname\csname
end#1\endcsname
 \renewenvironment{#1}%
   {\linenomath\csname old#1\endcsname}%
   {\csname oldend#1\endcsname\endlinenomath}%
}
\newcommand*\patchBothAmsMathEnvironmentsForLineno[1]{%
  \patchAmsMathEnvironmentForLineno{#1}%
  \patchAmsMathEnvironmentForLineno{#1*}%
}
\AtBeginDocument{%
\patchBothAmsMathEnvironmentsForLineno{equation}%
\patchBothAmsMathEnvironmentsForLineno{align}%
\patchBothAmsMathEnvironmentsForLineno{flalign}%
\patchBothAmsMathEnvironmentsForLineno{alignat}%
\patchBothAmsMathEnvironmentsForLineno{gather}%
\patchBothAmsMathEnvironmentsForLineno{multline}%
\patchBothAmsMathEnvironmentsForLineno{eqnarray}%
}

\usepackage{hyperxmp}

\usepackage[pdftex,
            pdfauthor={\paperauthors},
            pdftitle={\paperasciititle},
            pdfkeywords={\paperkeywords},
            pdfcopyright={Copyright (C) \papercopyright},
            pdflicenseurl={\paperlicenceurl}]{hyperref}

\usepackage[colorinlistoftodos,textsize=scriptsize]{todonotes}

\usepackage[all]{hypcap} 

\input{symbols/lhcb-symbols-def} 
\input{symbols/fixed-symbols-def}

\input{symbols/user-symbols-def}

\usepackage{cite} 
\usepackage{mciteplus}

\usepackage{subcaption}

%% file: symbols/lhcb-symbols-def.tex
\usepackage{xspace} 
\usepackage{upgreek}

\def\lhcb   {\mbox{LHCb}\xspace}

\def\MagUp {\mbox{\em Mag\kern -0.05em Up}\xspace}

\ifthenelse{\boolean{uprightparticles}}%
{

 \def\Ppi         {\ensuremath{\uppi}\xspace}                 
                  
 \def\Prho        {\ensuremath{\uprho}\xspace}

 \def\Ppsi        {\ensuremath{\uppsi}\xspace}

 \def\PDelta      {\ensuremath{\Delta}\xspace}                 
 \def\PXi         {\ensuremath{\Xi}\xspace}                 
 \def\PLambda     {\ensuremath{\Lambda}\xspace}                 
 \def\PSigma      {\ensuremath{\Sigma}\xspace}                 
 \def\POmega      {\ensuremath{\Omega}\xspace}                 
 \def\PUpsilon    {\ensuremath{\Upsilon}\xspace}

 \def\PB      {\ensuremath{\mathrm{B}}\xspace}                 
                  
 \def\PD      {\ensuremath{\mathrm{D}}\xspace}

 \def\PJ      {\ensuremath{\mathrm{J}}\xspace}                 
 \def\PK      {\ensuremath{\mathrm{K}}\xspace}

 \def\Pb      {\ensuremath{\mathrm{b}}\xspace}                 
 \def\Pc      {\ensuremath{\mathrm{c}}\xspace}

 \def\Pi      {\ensuremath{\mathrm{i}}\xspace}

 \def\Pp      {\ensuremath{\mathrm{p}}\xspace}

 \def\Ps      {\ensuremath{\mathrm{s}}\xspace}

 \def\thebaroffset{0.0em}
}
{

 \def\Ppi         {\ensuremath{\pi}\xspace}                 
                  
 \def\Prho        {\ensuremath{\rho}\xspace}

 \def\Ppsi        {\ensuremath{\psi}\xspace}                 
                  
 \mathchardef\PDelta="7101
 \mathchardef\PXi="7104
 \mathchardef\PLambda="7103
 \mathchardef\PSigma="7106
 \mathchardef\POmega="710A
 \mathchardef\PUpsilon="7107
                  
 \def\PB      {\ensuremath{B}\xspace}                 
                  
 \def\PD      {\ensuremath{D}\xspace}

 \def\PJ      {\ensuremath{J}\xspace}                 
 \def\PK      {\ensuremath{K}\xspace}

 \def\Pb      {\ensuremath{b}\xspace}                 
 \def\Pc      {\ensuremath{c}\xspace}

 \def\Pi      {\ensuremath{i}\xspace}

 \def\Pp      {\ensuremath{p}\xspace}

 \def\Ps      {\ensuremath{s}\xspace}

 \def\thebaroffset{0.18em}
}
\newcommand{\offsetoverline}[2][\thebaroffset]{\kern #1\overline{\kern -#1 #2}}%

\makeatletter
\ifcase \@ptsize \relax
  \newcommand{\miniscule}{\@setfontsize\miniscule{4}{5}}
\or
  \newcommand{\miniscule}{\@setfontsize\miniscule{5}{6}}
\or
  \newcommand{\miniscule}{\@setfontsize\miniscule{5}{6}}
\fi
\makeatother

\DeclareRobustCommand{\optbar}[1]{\shortstack{{\miniscule (\rule[.5ex]{1.25em}{.18mm})}
  \\ [-.7ex] $#1$}}

\def\squark    {{\ensuremath{\Ps}}\xspace}

\def\cquark    {{\ensuremath{\Pc}}\xspace}

\def\bquark    {{\ensuremath{\Pb}}\xspace}

\def\pion   {{\ensuremath{\Ppi}}\xspace}
\def\piz    {{\ensuremath{\pion^0}}\xspace}
\def\pip    {{\ensuremath{\pion^+}}\xspace}
\def\pim    {{\ensuremath{\pion^-}}\xspace}

\def\rhomeson {{\ensuremath{\Prho}}\xspace}

\def\rhop     {{\ensuremath{\rhomeson^+}}\xspace}

\def\kaon    {{\ensuremath{\PK}}\xspace}

\def\KorKbar {\kern \thebaroffset\optbar{\kern -\thebaroffset \PK}{}\xspace}

\def\Kp      {{\ensuremath{\kaon^+}}\xspace}
\def\Km      {{\ensuremath{\kaon^-}}\xspace}

\def\KS      {{\ensuremath{\kaon^0_{\mathrm{S}}}}\xspace}

\def\D       {{\ensuremath{\PD}}\xspace}

\def\DorDbar {\kern \thebaroffset\optbar{\kern -\thebaroffset \PD}\xspace}
\def\Dz      {{\ensuremath{\D^0}}\xspace}

\def\Dp      {{\ensuremath{\D^+}}\xspace}

\def\Ds      {{\ensuremath{\D^+_\squark}}\xspace}

\def\B       {{\ensuremath{\PB}}\xspace}

\def\BorBbar {\kern \thebaroffset\optbar{\kern -\thebaroffset \PB}\xspace}

\def\Bd      {{\ensuremath{\B^0}}\xspace}

\def\BdorBdbar {\kern \thebaroffset\optbar{\kern -\thebaroffset \Bd}\xspace}
\def\Bu      {{\ensuremath{\B^+}}\xspace}

\def\Bs      {{\ensuremath{\B^0_\squark}}\xspace}

\def\BsorBsbar {\kern \thebaroffset\optbar{\kern -\thebaroffset \Bs}\xspace}

\def\jpsi     {{\ensuremath{{\PJ\mskip -3mu/\mskip -2mu\Ppsi\mskip 2mu}}}\xspace}

\def\Y#1S{\ensuremath{\PUpsilon{(#1S)}}\xspace}

\def\proton      {{\ensuremath{\Pp}}\xspace}

\def\Lz          {{\ensuremath{\PLambda}}\xspace}

\def\LorLbar     {\kern \thebaroffset\optbar{\kern -\thebaroffset \PLambda}\xspace}

\def\Xires       {{\ensuremath{\PXi}}\xspace}

\def\Lc          {{\ensuremath{\Lz^+_\cquark}}\xspace}

\def\Xicp        {{\ensuremath{\Xires^+_\cquark}}\xspace}

\def\Xiccpp      {{\ensuremath{\Xires^{++}_{\cquark\cquark}}}\xspace}

\newcommand{\decay}[2]{\ensuremath{#1\!\to #2}\xspace}         

\def\to                 {\ensuremath{\rightarrow}\xspace}

\def\AT#1     {\ensuremath{A_{\mathrm{T}}^{#1}}\xspace}           

\def\C#1      {\ensuremath{\mathcal{C}_{#1}}\xspace}                       
\def\Cp#1     {\ensuremath{\mathcal{C}_{#1}^{'}}\xspace}                    
\def\Ceff#1   {\ensuremath{\mathcal{C}_{#1}^{\mathrm{(eff)}}}\xspace}        
\def\Cpeff#1  {\ensuremath{\mathcal{C}_{#1}^{'\mathrm{(eff)}}}\xspace}       
\def\Ope#1    {\ensuremath{\mathcal{O}_{#1}}\xspace}                       
\def\Opep#1   {\ensuremath{\mathcal{O}_{#1}^{'}}\xspace}                    

\newcommand{\nospaceunit}[1]{\ensuremath{\text{#1}}}       
\newcommand{\aunit}[1]{\ensuremath{\text{\,#1}}}       

\newcommand{\tev}{\aunit{Te\kern -0.1em V}\xspace}
\newcommand{\gev}{\aunit{Ge\kern -0.1em V}\xspace}
\newcommand{\mev}{\aunit{Me\kern -0.1em V}\xspace}
\newcommand{\kev}{\aunit{ke\kern -0.1em V}\xspace}
\newcommand{\ev}{\aunit{e\kern -0.1em V}\xspace}
\newcommand{\mevc}{\ensuremath{\aunit{Me\kern -0.1em V\!/}c}\xspace}
\newcommand{\gevc}{\ensuremath{\aunit{Ge\kern -0.1em V\!/}c}\xspace}
\newcommand{\mevcc}{\ensuremath{\aunit{Me\kern -0.1em V\!/}c^2}\xspace}
\newcommand{\gevcc}{\ensuremath{\aunit{Ge\kern -0.1em V\!/}c^2}\xspace}

\def\mum  {\ensuremath{\,\upmu\nospaceunit{m}}\xspace}

\def\fb   {\ensuremath{\aunit{fb}}\xspace}
\def\invfb   {\ensuremath{\fb^{-1}}\xspace}

\newcommand{\stat}{\aunit{(stat)}\xspace}
\newcommand{\syst}{\aunit{(syst)}\xspace}

\newcommand{\chisq}{\ensuremath{\chi^2}\xspace}

\newcommand{\chisqip}{\ensuremath{\chi^2_{\text{IP}}}\xspace}

\def\gsim{{~\raise.15em\hbox{$>$}\kern-.85em
          \lower.35em\hbox{$\sim$}~}\xspace}
\def\lsim{{~\raise.15em\hbox{$<$}\kern-.85em
          \lower.35em\hbox{$\sim$}~}\xspace}

\newcommand{\abs}[1]{\ensuremath{\left\|#1\right\|}} 

\def\sPlot{\mbox{\em sPlot}\xspace}

\def\pt         {\ensuremath{p_{\mathrm{T}}}\xspace}

\def\evtgen     {\mbox{\textsc{EvtGen}}\xspace}

\def\geant      {\mbox{\textsc{Geant4}}\xspace}

\def\photos     {\mbox{\textsc{Photos}}\xspace}

\def\pythia     {\mbox{\textsc{Pythia}}\xspace}

\xspace

\def\tell1  {TELL1\xspace}
\def\ukl1   {UKL1\xspace}



%% file: symbols/fixed-symbols-def.tex
\usepackage{xspace} 
\usepackage{upgreek}

\def\DeltaOrDeltabar  {\kern 0.18em\optbar{\kern -0.18em \PDelta}{}\xspace}
\def\XiOrXibar        {\kern 0.18em\optbar{\kern -0.18em \PXi}{}\xspace}
\def\SigmaOrSigmabar  {\kern 0.18em\optbar{\kern -0.18em \PSigma}{}\xspace}
\def\OmegaOrOmegabar  {\kern 0.18em\optbar{\kern -0.18em \POmega}{}\xspace}

\def\Lcp          {{\ensuremath{\Lz^+_\cquark}}\xspace}

\def\Xicp         {{\ensuremath{\Xires^+_\cquark}}\xspace}

\def\Xiccpp       {{\ensuremath{\Xires^{++}_{\cquark\cquark}}}\xspace}

\newcommand{\TeVnosp}{\ifthenelse{\boolean{inbibliography}}{\ensuremath{~T\kern -0.05em eV}}{\ensuremath{\mathrm{Te\kern -0.1em V}}}}
\newcommand{\GeVnosp}{\ensuremath{\mathrm{Ge\kern -0.1em V}}}
\newcommand{\MeVnosp}{\ensuremath{\mathrm{Me\kern -0.1em V}}}
\newcommand{\keVnosp}{\ensuremath{\mathrm{ke\kern -0.1em V}}}
\newcommand{\eVnosp}{\ensuremath{\mathrm{e\kern -0.1em V}}}
\newcommand{\GeVcnosp}{\ensuremath{{\mathrm{Ge\kern -0.1em V\!/}c}}}
\newcommand{\MeVcnosp}{\ensuremath{{\mathrm{Me\kern -0.1em V\!/}c}}}
\newcommand{\GeVccnosp}{\ensuremath{{\mathrm{Ge\kern -0.1em V\!/}c^2}}}
\newcommand{\MeVccnosp}{\ensuremath{{\mathrm{Me\kern -0.1em V\!/}c^2}}}
\newcommand{\GeVGeVccccnosp}{\ensuremath{{\mathrm{Ge\kern -0.1em V^2\!/}c^4}}}

\def\genxicctwo   {\mbox{\textsc{GenXicc2.0}}\xspace}

\renewcommand{\abs}[1]{\lvert#1\rvert}

\def\XiccppDecay{{\ensuremath{\Xiccpp\to\Lc\Km\pip\pip}}\xspace}
\def\XiccppToXicpiDecay{{\ensuremath{\Xiccpp\to\Xicp\pip}}\xspace}

\newcommand{\xx}{\ensuremath{\kern 0.5em }}

\def\sPlot{\mbox{\em sPlot}\xspace}

\usepackage{multirow} 
\usepackage{booktabs} 
\usepackage{rotating}

%% file: symbols/user-symbols-def.tex
\usepackage{xspace} 
\usepackage{upgreek}

%% file: mass-result-macros.tex
\def\preXiccppAveMass{\ensuremath{3621.24\,\pm 0.65\,{\rm(stat)\,}
	 \pm 0.31\,{\rm(syst)}}\mevcc}

\def\XiccppMassB{\mbox{\ensuremath{3621.55\,\pm 0.23\,{\rm(stat)\,}
	 \pm 0.30\,{\rm(syst)}} \mevcc}}

%% file: title-LHCb-PAPER.tex

\begin{titlepage}
\pagenumbering{roman}

\vspace*{-1.5cm}
\centerline{\large EUROPEAN ORGANIZATION FOR NUCLEAR RESEARCH (CERN)}
\vspace*{1.5cm}
\noindent
\begin{tabular*}{\linewidth}{lc@{\extracolsep{\fill}}r@{\extracolsep{0pt}}}
\ifthenelse{\boolean{pdflatex}}
{\vspace*{-1.5cm}\mbox{\!\!\!\includegraphics[width=.14\textwidth]{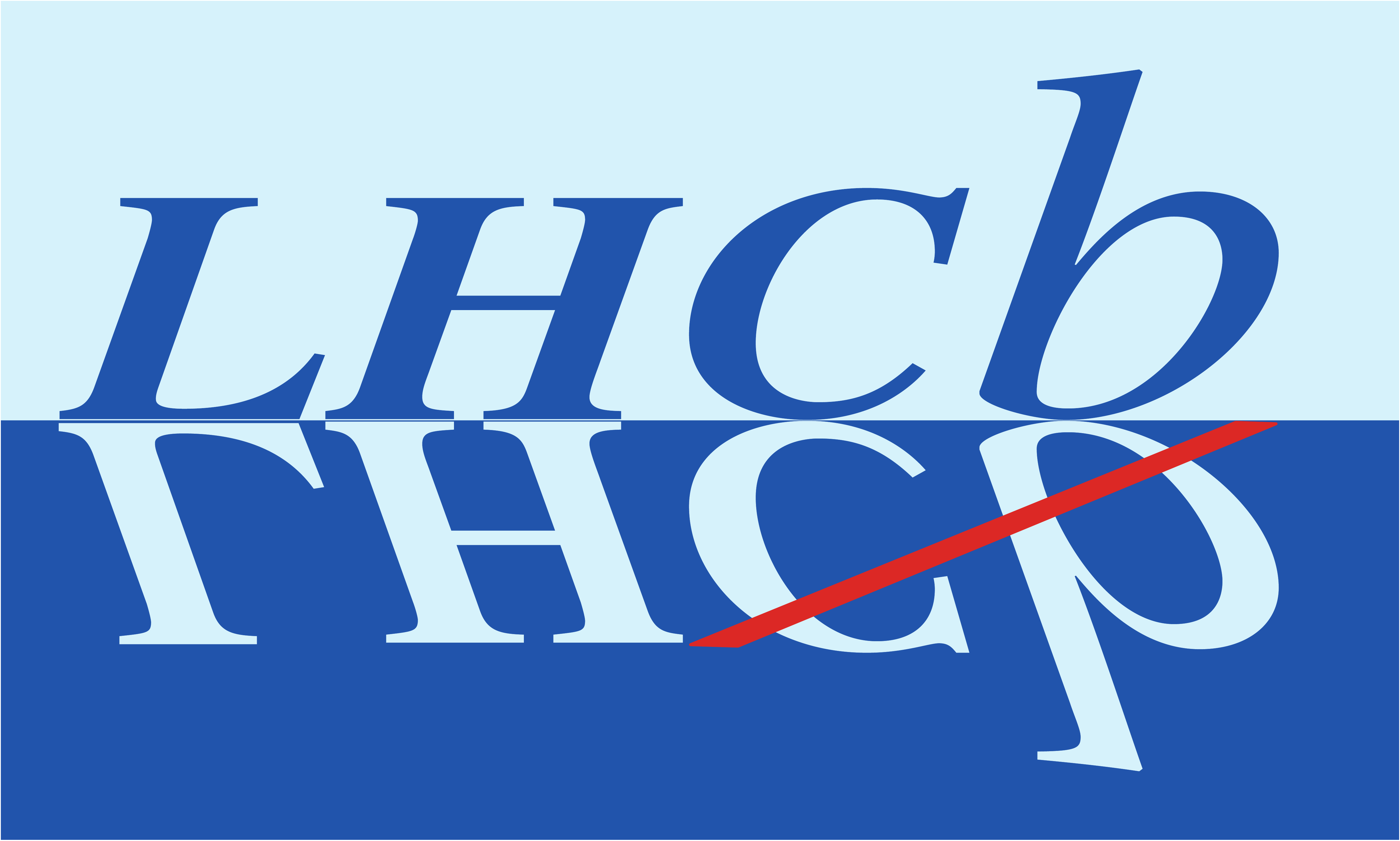}} & &}%
{\vspace*{-1.2cm}\mbox{\!\!\!\includegraphics[width=.12\textwidth]{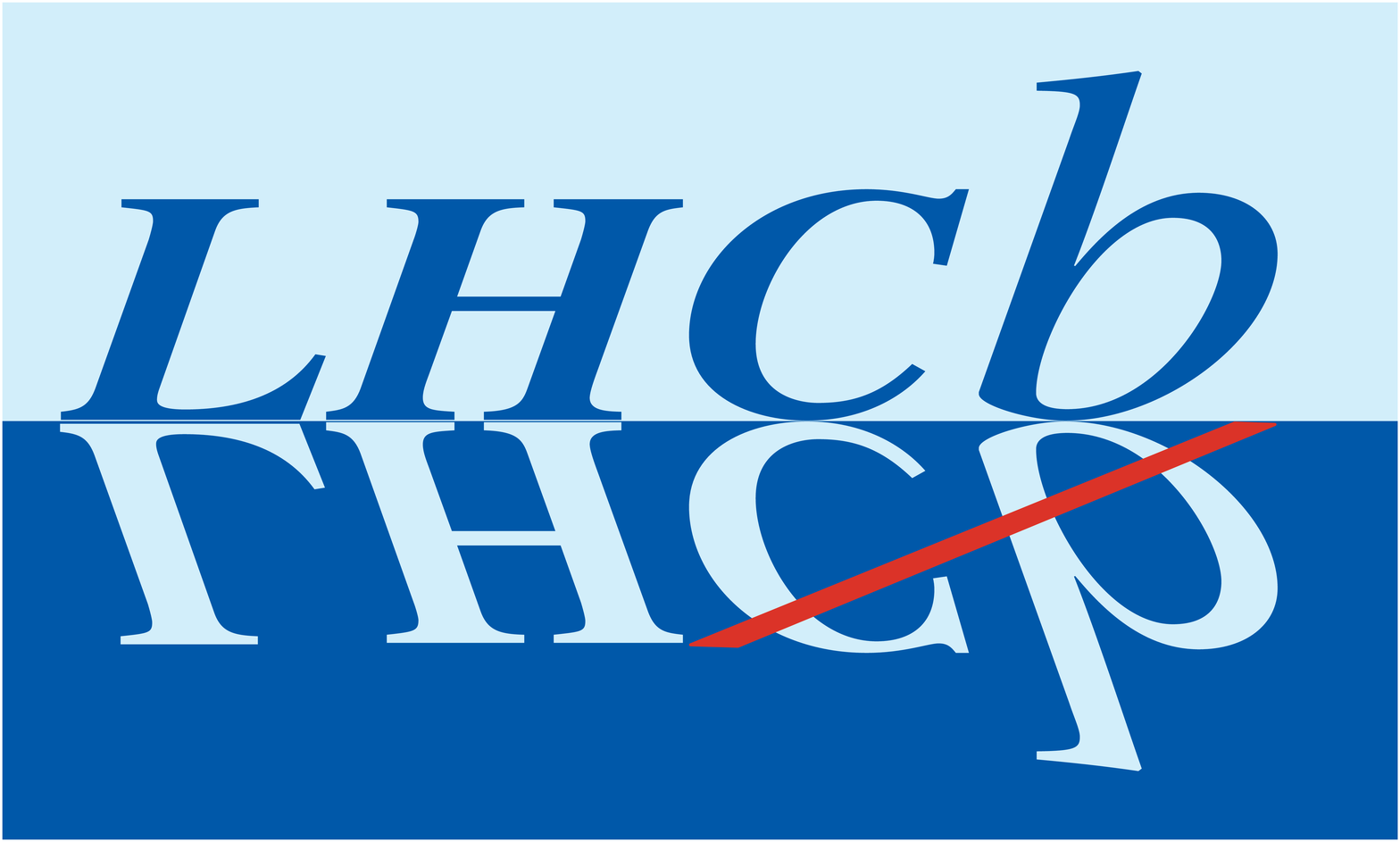}} & &}%
\\
 & & CERN-EP-2019-258 \\  
 & & LHCb-PAPER-2019-037 \\  
 & & February 11, 2020 \\ 
\end{tabular*}

\vspace*{4.0cm}

{\normalfont\bfseries\boldmath\huge
\begin{center}
\papertitle 
\end{center}
}

\vspace*{2.0cm}

\begin{center}
\paperauthors\footnote{Authors are listed at the end of this paper.}
\end{center}

\vspace{\fill}

\begin{abstract}

  \noindent
  A measurement of the \Xiccpp mass is performed using data collected by the
  \lhcb experiment between 2016 and 2018 in
  \proton\proton collisions at a centre-of-mass energy of 13\tev, corresponding to an integrated luminosity of $5.6\invfb$.
  The $\Xiccpp$ candidates are reconstructed via the decay modes
  \mbox{\decay{\Xiccpp}{\Lcp\Km\pip\pip}} and \XiccppToXicpiDecay. 
   The result, \XiccppMassB, is the most precise measurement of the $\Xiccpp$ mass to date.

\end{abstract}

\vspace*{2.0cm}

\begin{center}
Published in JHEP 02 (2020) 049 
\end{center}

\vspace{\fill}

{\footnotesize 
\centerline{\copyright~\papercopyright. \href{\paperlicenceurl}{\paperlicence}.}}
\vspace*{2mm}

\end{titlepage}


\newpage
\setcounter{page}{2}
\mbox{~}
%
%
%
%

\cleardoublepage

%% file: mass-paper-body.tex
\section{Introduction}

Baryons containing two charm quarks and a lighter quark are predicted by the 
quark model~\cite{GellMann:1964nj,
Zweig:352337,*Zweig:570209} and provide an ideal system to study the dynamics of bound states of quarks.
Observation of the $\Xiccpp$ ($ccu$) baryon via decays to $\Lcp\Km\pip\pip$ and $\Xicp\pip$ final states 
has been reported 
by the \lhcb collaboration~\cite{LHCb-PAPER-2017-018, LHCb-PAPER-2018-026}.\footnote{The inclusion of charge conjugate modes is implied throughout this paper.} The $\Xiccpp$ mass 
was measured by the \lhcb collaboration to be $3621.24 \pm 0.65\stat \pm 0.31 \syst \mevcc$. 
Before the \lhcb observation, 
theoretical calculations using quark models\mbox{\cite{Ebert:2002ig, Roberts:2007ni, Karliner:2014gca}}, bag models~\cite{He:2004px}, the Faddeev method~\cite{Valcarce:2008dr}, quantum chromodynamics (QCD) sum rules\cite{wang2010analysis,Kiselev:2001fw,Zhang:2008rt, Aliev:2012ru}, potential models~\cite{Richard:2005jz} and lattice QCD~\cite{Lewis:2001iz, Flynn:2003vz,Liu:2009jc} predicted the mass of
this state  in the range 3450--3750\mevcc.  Most of the predictions using quark models are around 3600\mevcc while other methods have a larger spread. Theoretical calculations of the $\Xiccpp$ mass~\cite{ Weng:2018prd, Li:2019ekr, Wang:2018lhz, Qi:2017prd} after the LHCb observation
fall into a $\pm20\mevcc$ window around the experimental value measured by LHCb.

At present, the experimental uncertainty on the \Xiccpp mass is still large compared to that of the singly charmed baryons. This paper presents an updated measurement of the \Xiccpp mass using the decay modes \mbox{$\Xiccpp\to\Lcp(\to\proton\Km\pip)\Km\pip\pip$} and \mbox{$\Xiccpp\to\Xicp(\to\proton\Km\pip)\pip$}. The analysis uses a data sample corresponding to an integrated luminosity of 
5.6\invfb, 
collected by the LHCb experiment during 2016--2018 in \proton\proton collisions at a centre-of-mass energy of 13\tev. 
This measurement supersedes the results reported on the \Xiccpp mass in Refs.~\cite{LHCb-PAPER-2017-018, LHCb-PAPER-2018-026}, which only use $pp$ collision data at 13\tev taken in 2016, corresponding to an integrated luminosity of 1.7\invfb.
 
\section{Detector and simulation}
The \lhcb detector~\cite{Alves:2008zz,LHCb-DP-2014-002} is a single-arm forward spectrometer covering the pseudorapidity range $2<\eta <5$, designed for the study of particles containing \bquark or \cquark quarks. 
The detector includes a high-precision tracking system consisting of a silicon-strip vertex detector surrounding the $pp$ interaction region~\cite{LHCb-DP-2014-001}, a large-area silicon-strip detector located upstream of a dipole magnet with a bending power of about $4{\mathrm{\,Tm}}$, and three stations of silicon-strip detectors and straw drift tubes~\cite{LHCb-DP-2013-003, LHCb-DP-2017-001} placed downstream of the magnet.
The tracking system provides a measurement of the momentum of charged particles with a relative uncertainty that varies from 0.5\% at low momentum to 1.0\% at 200\gevc.   The momentum scale is calibrated using samples of $\Bu \rightarrow \jpsi K^+$ and $\jpsi \rightarrow \mu^+
\mu^-$ decays collected concurrently
with the data sample used for this analysis 
\cite{LHCb-PAPER-2012-048,LHCb-PAPER-2013-011}. The relative accuracy of 
this
procedure is estimated to be $3 \times 10^{-4}$ using samples of other
fully reconstructed $\bquark$-hadron, $\PUpsilon$ and
$\KS$ decays.
The minimum distance of a track to a primary $pp$ collision vertex (PV), the impact parameter (IP), is measured with a resolution of $(15+29/(\pt/\gevc))\mum$, where \pt is the 
momentum component transverse to the beam axis. 
Different types of charged hadrons are distinguished using information
from two ring-imaging Cherenkov detectors~\cite{LHCb-DP-2012-003}. 
Photons, electrons and hadrons are identified by a calorimeter system consisting of
scintillating-pad and preshower detectors, an electromagnetic
and a hadronic calorimeter. Muons are identified by a
system composed of alternating layers of iron and multiwire
proportional chambers~\cite{LHCb-DP-2012-002}.

The online event selection is performed by a trigger~\cite{LHCb-DP-2012-004}, 
which consists of a hardware stage, based on information from the calorimeter and muon
systems, followed by a software stage, which applies a full event
reconstruction. In between the two software stages, an alignment and calibration of the detector is performed in near real-time~\cite{LHCb-PROC-2015-011}.
This process allows the reconstruction of the \Xiccpp decays to be performed entirely in the software trigger, whose output is used as input to the present analysis.

Simulated samples are used to model the effects of the detector acceptance, optimise selections and verify the validity of the methods used in the measurement. In the simulation, \proton\proton collisions are generated using \pythia 8~\cite{Sjostrand:2007gs} with a \lhcb specific configuration~\cite{LHCb-PROC-2010-056}. 
The production of doubly charmed $\Xiccpp$ baryons is simulated using the dedicated generator \genxicctwo~\cite{Chang:2009va, *Wang:2012vj}. 
Decays of hadrons are described by \evtgen~\cite{Lange:2001uf}, in which final-state radiation is generated using \photos~\cite{Golonka:2005pn}.
The interaction of the generated particles with the detector, and its response, are implemented using the \geant toolkit~\cite{Allison:2006ve, *Agostinelli:2002hh} as described in Ref.~\cite{LHCb-PROC-2011-006}.
Sources of background, such as
those from $\Xiccpp\to{\it \Xi^{'+}_c}(\to\Xicp\gamma)\pip$ and $\Xiccpp\to\Xicp\rhop(\to\pip\piz)$, are studied using the
fast simulation package \mbox{\textsc{RapidSim}}~\cite{Cowan:2016tnm}.

\section{Event selection}
\label{sec:selection}
The reconstruction of \mbox{$\Xiccpp\to\Lcp(\to\proton\Km\pip)\Km\pip\pip$} and \mbox{$\Xiccpp\to\Xicp(\to\proton\Km\pip)\pip$} decays  
is similar to that used in 
 previous \lhcb analyses~\cite{LHCb-PAPER-2017-018, LHCb-PAPER-2018-026}.
Candidate \mbox{$\Xicp(\Lcp)\to p K^- \pip$} decays 
are reconstructed from three charged particles
identified as a $p$, $K^-$ and $\pip$ using information from the RICH  detectors~\cite{LHCb-DP-2012-003}. The charged particles are required to  
form a good-quality vertex and be inconsistent with originating from any PV.
The $ \Xicp (\Lcp)$ candidate is then combined with 
one  (three) additional charged particle(s) 
to form a \XiccppToXicpiDecay (\XiccppDecay) decay candidate. 
These additional particles must form a good-quality vertex with the $\Xicp (\Lcp)$ candidate, which is required to be upstream 
of the $\Xicp (\Lcp)$ decay vertex. 
Each \Xiccpp candidate is required to have $\pt>2\gevc$ and to
be consistent with originating from its associated PV. 
The associated PV is that with respect to which the $\Xiccpp$ candidate has the smallest $\chisqip$. The \chisqip is defined as the difference in \chisq of the PV fit with and without the particle in question.
To avoid candidates including duplicated tracks, 
each track pair is required to have an opening angle larger than 0.5 mrad or a momentum difference larger than 5\% of
the minimum momentum of the track pair.

In order to improve the signal purity, multivariate classifiers are trained to separate signal from background. The choice of classifier algorithms is based on their performance for each decay mode. A classifier based on the Boosted Decision Tree (BDT)  algorithm~\cite{Breiman,AdaBoost} implemented in the TMVA
toolkit~\cite{Hocker:2007ht,*TMVA4} is used for the the \XiccppDecay mode, while 
a Multilayer Perceptron (MLP) algorithm~\cite{Hocker:2007ht,*TMVA4} is used for the \XiccppToXicpiDecay mode.
The BDT for the \mbox{\XiccppDecay} decay is trained with simulated signal events as a signal proxy and wrong-sign \Lcp\Km\pip\pim combinations (3525--3725\mevcc) in data, where the two final-state pions have opposite charges, as a background proxy. Both the signal and  background proxies are required to pass the selection described above. Variables associated with the \Xiccpp candidates used in the training include 
the vertex-fit quality,
the \chisqip,
the angle between the momentum and the displacement vector,
the flight-distance \chisq between the PV and the decay vertex. 
The flight-distance $\chisq$ is defined as the $\chisq$ of the hypothesis that 
the decay vertex of the candidate coincides with its associated PV.
Variables associated with the decay products  of the \Xiccpp candidates (the $\Lcp$, $\Km$ and $\pip$) used in the training include their \pt and  \chisqip,
the vertex-fit quality of the \Lcp candidates and the smallest \pt among the \Lcp decay products ($\proton$, $\Km$ and $\pip$). 
Particle-identification information for the final-state particles is also used.

The threshold applied to the classifier response is determined by maximising the signal significance $S/\sqrt{S+B}$, 
where $S$ is the expected signal yield estimated using simulation,
and $B$ is the background yield evaluated in the upper sideband of data \mbox{(3800--3900\mevcc)} extrapolated to the signal region (3607--3635\mevcc). 
The multivariate classifier
for the \mbox{\XiccppToXicpiDecay} decay is developed following the same strategy as that for the \mbox{\XiccppDecay} decay.

After the full selection, an event may still contain more than one $\Xiccpp$ candidate.
According to studies on simulated decays and the wrong-sign control sample, multiple candidates in the same event may form a peaking structure 
in the mass distribution of the $\Xiccpp$ candidates 
if they are obtained from the same final-state
tracks, but via swapping two final state tracks (e.g. the \Km from the \Lc decay and 
the \Km from the \Xiccpp decay).
In this case,
one candidate is chosen randomly.
Other kinds of multiple candidates, which account for 8\% ($<1$\%) of the 
\XiccppDecay (\XiccppToXicpiDecay) signal events, 
are not removed since they do not form a peaking background.

\section{Mass measurement}
To improve the mass resolution, the invariant mass 
of a $\Xiccpp$ candidate is computed as
\begin{equation}
    \label{eq:dm}
\begin{aligned}
    m_{\rm cand}(\Xiccpp) &= m(\Lcp\Km\pip\pip)-m(\Lcp)+M_{\rm PDG}(\Lcp), \\
    m_{\rm cand}(\Xiccpp) &= m(\Xicp\pip)-m(\Xicp)+M_{\rm LHCb}(\Xicp),
\end{aligned}
\end{equation}
where $m(\Lcp\Km\pip\pip)$ and $m(\Xicp\pip)$ are the reconstructed $\Xiccpp$ masses; $m(\Lcp)$ and $m(\Xicp)$ are the reconstructed $\Lcp$ and $\Xicp$ masses; $M_{\rm PDG}(\Lcp)$ is the known value of the \Lcp mass; $M_{\rm LHCb}(\Xicp)$ 
is the known value of the \Xicp mass. 
The known value of the $\Lcp$
mass is $2286.46 \pm 0.14 \mevcc$\cite{Aubert:2005gt, PDG2019},
and that of the $\Xicp$ mass 
is determined to be $2467.97 \pm 0.22 \mevcc$ using $M_{\rm PDG}(\Lcp)$ and the difference between $m(\Xicp)$ and $m(\Lcp)$ of $181.51 \pm 0.14 \pm 0.10 \mevcc$ measured by the LHCb collaboration~\cite{LHCb-PAPER-2014-021}.

\begin{figure}
\centering
\includegraphics[width=0.48\linewidth]{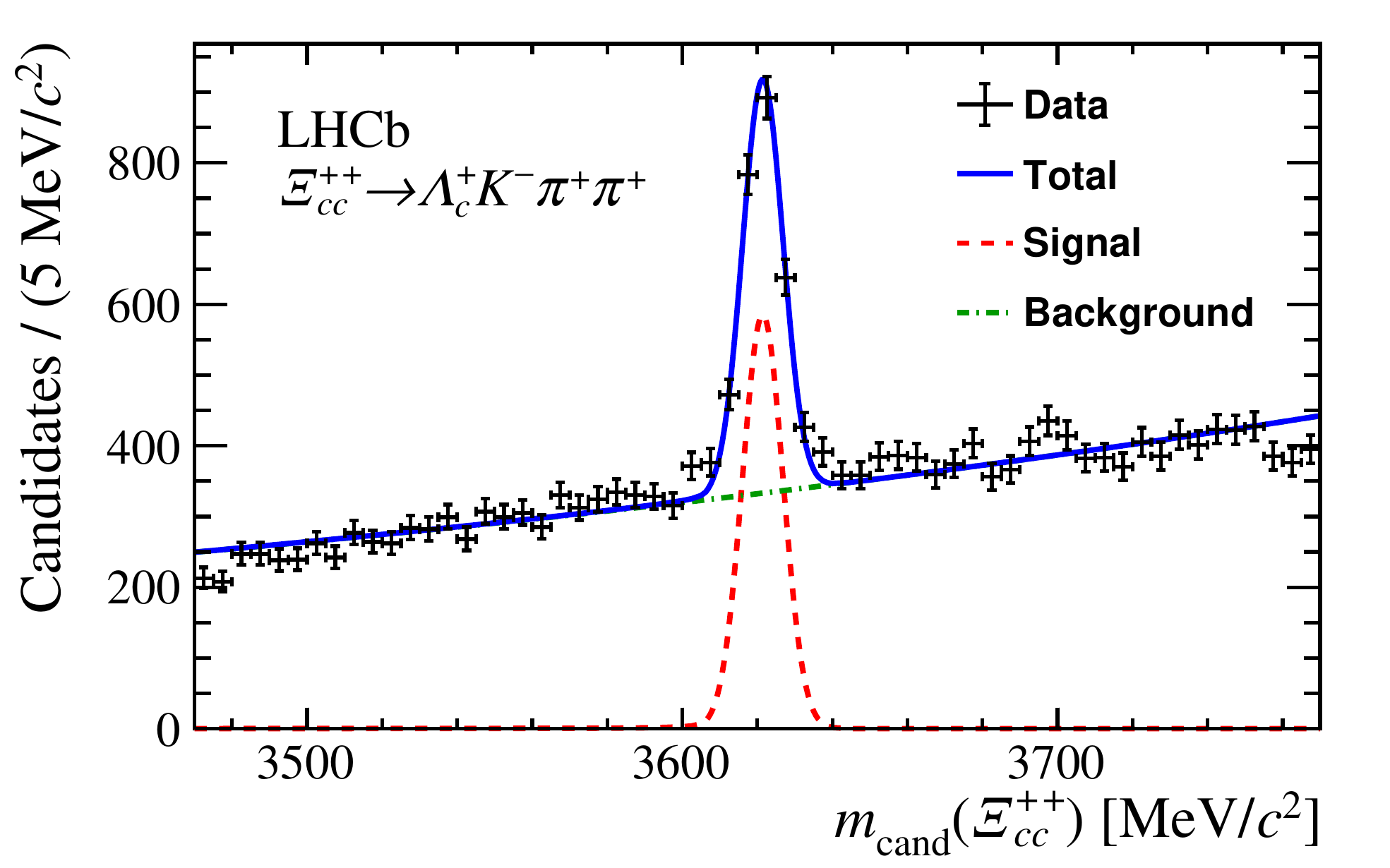}
\includegraphics[width=0.48\linewidth]{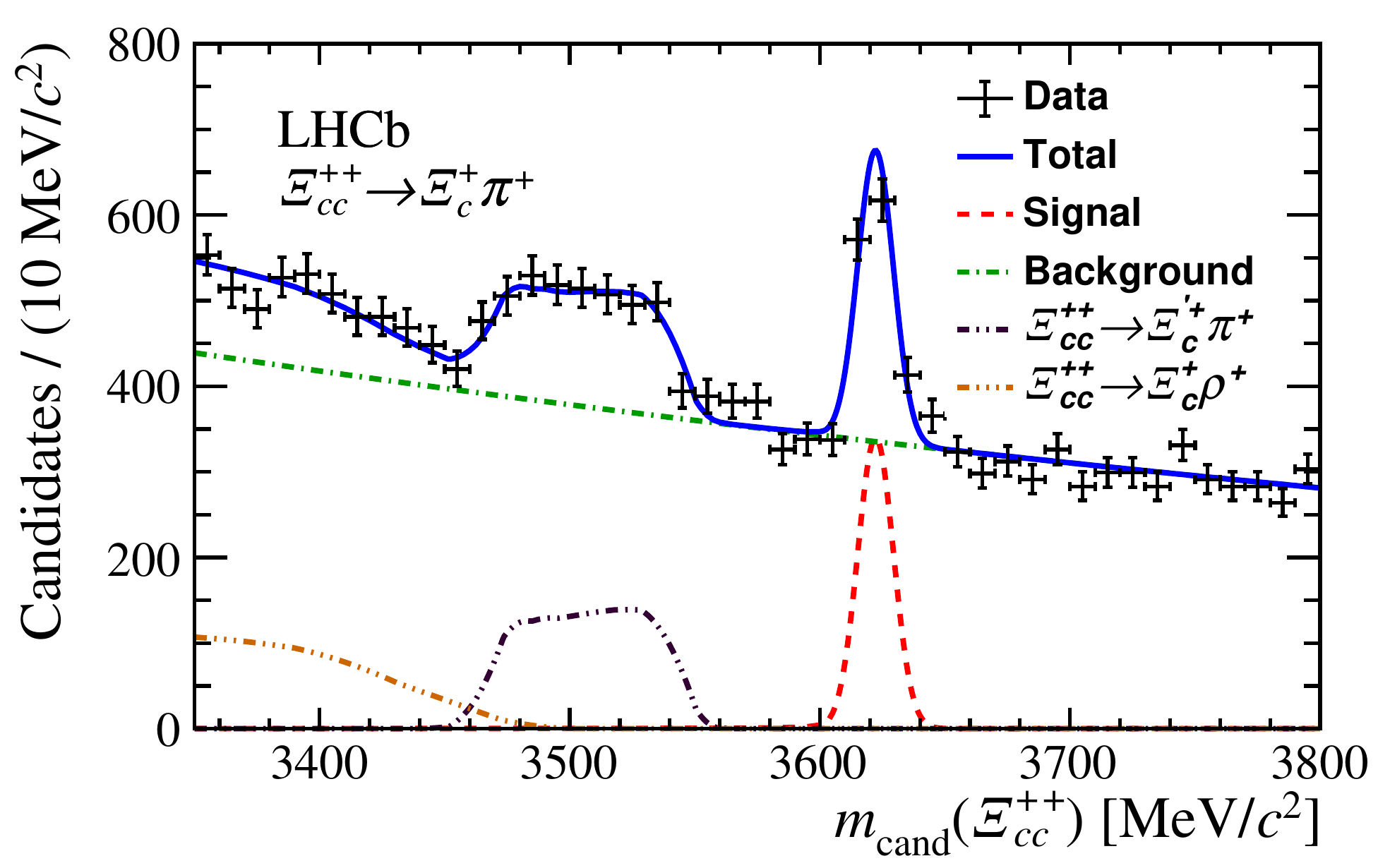}
\caption{Invariant-mass distributions of (left) \XiccppDecay and (right) \XiccppToXicpiDecay decay candidates, in the mass range of  3470--3770\mevcc and 3350--3800\mevcc, respectively. The results of unbinned extended maximum-likelihood fits to the mass distributions are indicated by the  blue solid lines.}
\label{fig:invariant-mass}
\end{figure}

The $m_{\rm cand}(\Xiccpp)$ mass distributions of the selected $\Xiccpp$ candidates are shown in Fig.~\ref{fig:invariant-mass} for the $\Xiccpp\to\Lcp\Km\pip\pip$ and \XiccppToXicpiDecay  decay modes. The $\Xiccpp$ mass is determined by performing unbinned extended maximum-likelihood fits to the two $m_{\rm cand}(\Xiccpp)$ mass distributions. 
The signal components are described by a modified Gaussian function with a power-law tail on the left-hand side of the distribution~\cite{Skwarnicki:1986xj}, parameterised as
\begin{equation}\label{eq:crystall-ball-function}
f(x; \alpha, N, \bar{x}, \sigma)=
\begin{cases}
 e^{-\frac{1}{2}(\frac{x-\bar{x}}{\sigma})^2} &{\rm for\,} \frac{x-\bar{x}}{\sigma}>-\alpha\\
 \left(\frac{N}{\abs{\alpha}}\right)^N e^{-\frac{\abs{\alpha}^2}{2}} \left(\frac{N}{\abs{\alpha}}-\abs{\alpha}-\frac{x-\bar{x}}{\sigma}\right)^{-N}  &{\rm for\,} \frac{x-\bar{x}}{\sigma}\leq-\alpha.
\end{cases}
\end{equation}
The peak position, $\overline{x}$, and width, $\sigma$,  of the function are allowed to vary in the fit. The power-law tail parameters, $N$ and $\alpha$, are fixed from simulation. The background from randomly associated tracks is modelled using  
an exponential function. 
Background contributions from the partially reconstructed decays $\Xiccpp\to{\it \Xi^{'+}_c}(\to\Xicp\gamma)\pip$ and $\Xiccpp\to\Xicp\rhop(\to\pip\piz)$, where photons and neutral $\piz$ mesons  are not reconstructed, can contribute to the \XiccppToXicpiDecay decay mode. The mass line shapes of these partially reconstructed backgrounds are determined from simulation.

The fits return signal yields of
$1598 \pm 64$ 
and 
$616 \pm 47$
for the $\XiccppDecay$ 
and 
$\XiccppToXicpiDecay$ decay
modes, respectively. 
The peak positions are determined with the $\XiccppDecay$ and 
$\XiccppToXicpiDecay$ decay modes to be 
$3622.08 \pm 0.24\mevcc$
and 
$3622.37 \pm 0.60\mevcc$, respectively, where the uncertainty is statistical only. 

Due to multiple scattering, the opening angle between the $\Xiccpp$ decay products can be increased or decreased. This can bias both the resulting \Xiccpp mass and the measured decay length. Since the selection is more efficient for candidates with larger reconstructed decay lengths, and the decay length is correlated with the mass by the effect of the multiple scattering, this can bias the $\Xiccpp$ mass measurement.
This effect was studied with charmed hadrons, $\Dp, \Dz, \Ds, \Lcp$, and was found to be well reproduced by simulation~\cite{LHCb-PAPER-2017-018}. 
Corresponding corrections of $-0.61\pm0.09\mevcc$ for \XiccppDecay and $-0.45\pm0.09\mevcc$ for \XiccppToXicpiDecay are determined using simulated candidates
by comparing the fitted mass with signal candidates before and after applying the event selection. These corrections are applied to the fitted mass values. The uncertainties are due to the limited size of simulated samples,
and are taken as the systematic uncertainties from the selection-induced bias on the $\Xiccpp$ mass.
The difference 
of the kinematic distributions 
in simulation and data is considered as a systematic uncertainty and is discussed in Sec.~\ref{sec:sys_uncertainties}.

Low-momentum photons emitted by the final-state particles are not reconstructed. This distorts the reconstructed mass distribution and can bias the fitted mass value. 
This effect is studied with the simulation. 
To disentangle this effect from those due to reconstruction, 
the mass of the \Xiccpp candidates calculated with the true momenta of the final-state particles is
smeared with different resolution values. 
The difference between the fitted and input mass values is studied as a function of the mass resolution, and
the difference corresponding to the mass resolution in data is taken as a correction. Alternative signal models are also considered. The largest difference of the fitted mass with final-state radiation corrections between the nominal and the alternative is quoted as the uncertainty. Following the procedure described above, the corrections due to the final-state radiation are determined as $0.06\pm0.05\mevcc$ and
$0.03 \pm 0.16\mevcc$ for the \XiccppDecay and \XiccppToXicpiDecay decay modes, respectively. The uncertainties 
on the corrections are due to the limited size of the simulated samples, and the difference between the corrections with different signal models. 

\section{Systematic uncertainties}
\label{sec:sys_uncertainties}

The dominant source of systematic uncertainty on the mass measurement  
is due to the 
momentum-scale  calibration~\cite{LHCb-PAPER-2012-048,LHCb-PAPER-2013-011}.  It amounts to 0.21\mevcc for the \XiccppDecay decay, and 0.34\mevcc for the \XiccppToXicpiDecay decay due to larger $Q$-value. A further uncertainty arises from the correction for energy
loss in the spectrometer, which is known with 10\% accuracy \cite{LHCb-DP-2014-002}.
This uncertainty was studied in Ref.~\cite{LHCb-PAPER-2013-011}, and amounts to 0.03\mevcc for \mbox{\Dz\to\Kp\Km\pip\pim} decays. The uncertainties on the \Xiccpp mass are scaled from that of the \Dz decay by the number of final-state particles, and are determined to be 0.05\mevcc and 0.03\mevcc for the \XiccppDecay and \XiccppToXicpiDecay decays, respectively.

Differences between kinematic distributions in simulation and data are treated as sources of systematic uncertainties on the  corrections due to the selection procedure. 
The kinematic variables used in the event selection that are found to affect the corrections are listed below: 
\pt of \Xiccpp candidates; 
the angle between the momentum and the displacement vector from the PV to the decay vertex of the \Xiccpp and \Lcp (\Xicp) candidates; the \chisqip of \Xiccpp and \Lcp (\Xicp) candidates and their decay products; the BDT (MLP) response; and the particle identification information. 
The distributions of these variables in simulation are weighted to match those in data where the background is subtracted by means of the \sPlot technique~\cite{Pivk:2004ty}. Then, the corrections obtained with the weights are compared to those without weights, 
and largest variations of the corrections
are taken as systematic uncertainties, 
which are 
0.09\mevcc
and 0.05\mevcc
for the \XiccppDecay and  \XiccppToXicpiDecay decays, respectively.

The uncertainty related to the background description is estimated by repeating the fits with alternative models which include first and second-order polynomial functions.
For the \XiccppDecay decay, the fit with a second-order polynomial function has better fit quality, but returns  identical fitted mass. 
The largest changes on the fitted mass value are found to be 0.01\mevcc 
and 0.04\mevcc
for the \XiccppDecay and \XiccppToXicpiDecay decays, respectively, which are assigned as systematic uncertainties.

The mass of \Xiccpp candidates also depends on the value of the known \Lcp and \Xicp masses. The uncertainties on the \Lcp mass and on the mass difference between the \Xicp and \Lcp are propagated to the \Xiccpp mass measurement. The corresponding uncertainties on the \Xiccpp mass are  $0.14\mevcc$ and $0.22\mevcc$ for the \XiccppDecay decay and the \XiccppToXicpiDecay decay, respectively.  

The sources of systematic uncertainty considered in this analysis are summarised in Table~\ref{tab:syst}. 
When computing the total uncertainty, the uncertainty on the momentum-scale calibration of the $\Xicp$ mass from Ref.~\cite{LHCb-PAPER-2014-021} is assumed to be fully correlated to that of the \Xiccpp mass. The total systematic uncertainty is calculated by summing the individual sources of uncertainty in quadrature.

\begin{table}
\centering
\caption{Systematic uncertainties on the \Xiccpp mass measurements using \XiccppDecay and \XiccppToXicpiDecay decays. The total systematic uncertainty on each mode is calculated by summing the individual sources of uncertainty in quadrature, except for the uncertainty on the momentum-scale calibration of the $\Xicp$ mass~\cite{LHCb-PAPER-2014-021}, that is added linearly to that of the \Xiccpp mass.}
\begin{tabular}{l|c|c}
&\multicolumn{2}{c}{Uncertainty~$[\mevcc]$}\\ \hline
Source & \XiccppDecay & \XiccppToXicpiDecay \\ \hline
   
Momentum-scale calibration  & 0.21         & 0.34 \\
Energy-loss correction  & 0.05        & 0.03 \\
Simulation/data agreement  & 0.09                 & 0.05 \\
Selection-induced bias on the \Xiccpp mass  & 0.09   & 0.09 \\
Final-state radiation  & 0.05       & 0.16 \\
Background model  & 0.01                 & 0.04 \\
\Lcp, \Xicp mass  & 0.14              & 0.22 \\ \hline
Total     & 0.29  & 0.49 \\ 

\end{tabular}
\label{tab:syst}
\end{table}

\section{Results and summary}
\label{sec:summary}
\begin{figure}[b]
\centering
\includegraphics[width=0.6\linewidth]{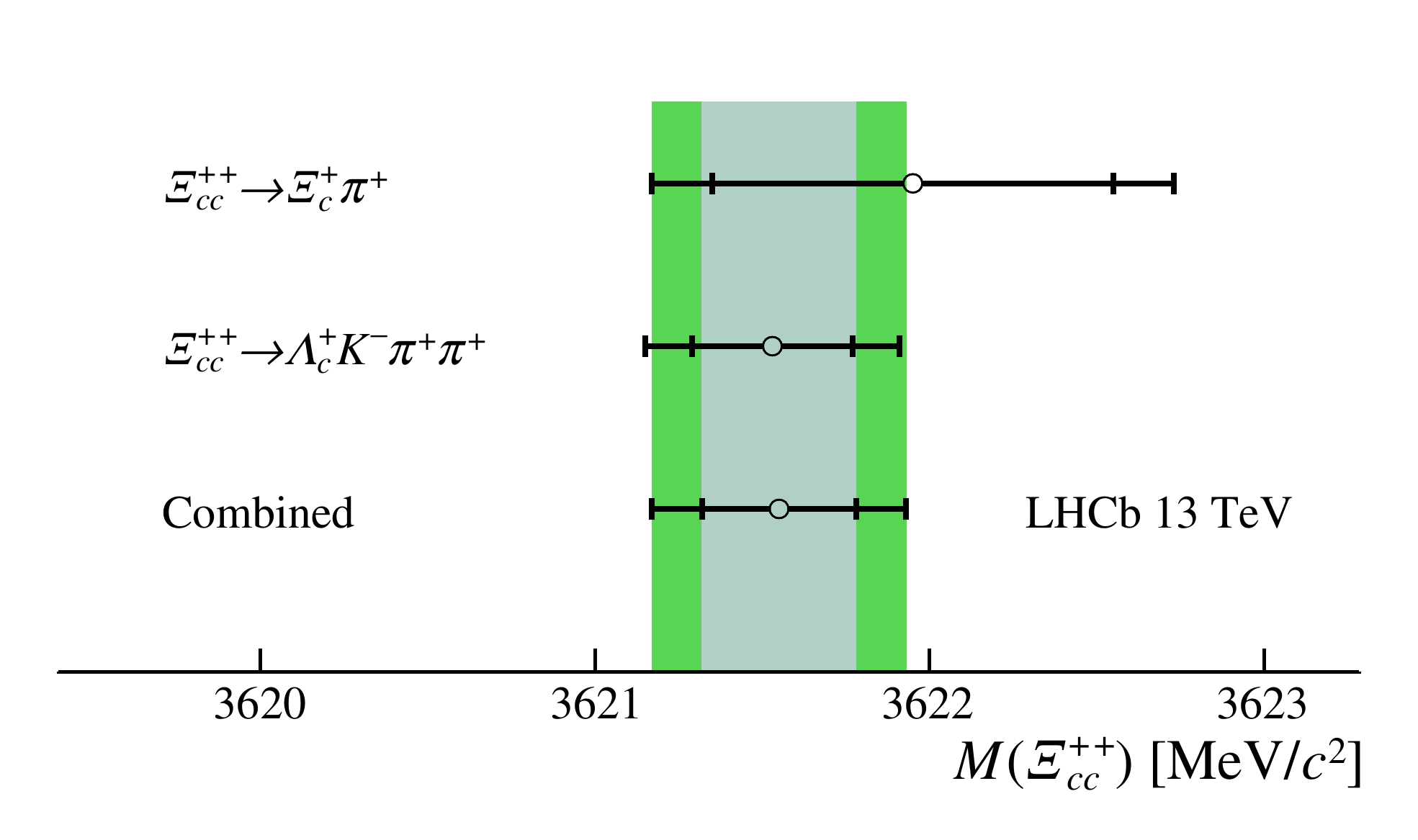}
\caption{Measured $\Xiccpp$ mass values and uncertainties obtained with the decay modes \mbox{\XiccppDecay} and \mbox{\XiccppToXicpiDecay}. The combination is performed using the best linear unbiased estimator~\cite{Lyons:1988rp, Valassi:2003mu}. The inner error bars represent statistical uncertainties and the outer error bars represent the quadratic sum of statistical and systematic uncertainties. 
The inner and outer green bands correspond to the uncertainties on the averaged value.}
\label{fig:BLUE}
\end{figure}

The resulting values of the $\Xiccpp$ mass 
using the  \mbox{\XiccppDecay} and \mbox{\XiccppToXicpiDecay} decay modes are 
$3621.53 \pm 0.24 \pm 0.29 \mevcc$, 
and 
$3621.95 \pm 0.60 \pm 0.49 \mevcc$, respectively, 
including corrections and systematic uncertainties. 
The combination of the two measurements is performed using 
the Best Linear Unbiased Estimator (BLUE)~\cite{Lyons:1988rp, Valassi:2003mu}. 
The combined \Xiccpp mass is determined to be 
 \begin{align*}
 \XiccppMassB.
  \end{align*}
In the combination, the correlation between the $\Lcp$ and $\Xicp$ masses~\cite{LHCb-PAPER-2014-021, PDG2019} is taken into account. 
Uncertainties arising from the momentum-scale calibration, energy-loss corrections, and final-state radiation
are assumed to be 100\% correlated while other sources of systematic uncertainty are assumed to be  uncorrelated.
The individual mass measurements and the resulting combination are illustrated in Fig.~\ref{fig:BLUE}.
The averaged mass is dominated by the result for the \XiccppDecay mode, due to its larger yield and smaller momentum-scale uncertainty relative to that of the \XiccppToXicpiDecay decay. This is the most precise measurement of the $\Xiccpp$ mass to date, improving upon the previous weighted average mass value of \preXiccppAveMass from Ref.~\cite{LHCb-PAPER-2018-026}.

\section*{Acknowledgements}

\noindent We thank Chao-Hsi Chang, Cai-Dian L\"u, Xing-Gang Wu, and Fu-Sheng Yu for the discussions on the production and decays of double-heavy-flavor baryons. We express our gratitude to our colleagues in the CERN
accelerator departments for the excellent performance of the LHC. We thank the technical and administrative staff at the LHCb
institutes.
We acknowledge support from CERN and from the national agencies:
CAPES, CNPq, FAPERJ and FINEP (Brazil);
MOST and NSFC (China);
CNRS/IN2P3 (France);
BMBF, DFG and MPG (Germany);
INFN (Italy);
NWO (Netherlands);
MNiSW and NCN (Poland);
MEN/IFA (Romania);
MSHE (Russia);
MinECo (Spain);
SNSF and SER (Switzerland);
NASU (Ukraine);
STFC (United Kingdom);
DOE NP and NSF (USA).
We acknowledge the computing resources that are provided by CERN, IN2P3
(France), KIT and DESY (Germany), INFN (Italy), SURF (Netherlands),
PIC (Spain), GridPP (United Kingdom), RRCKI and Yandex
LLC (Russia), CSCS (Switzerland), IFIN-HH (Romania), CBPF (Brazil),
PL-GRID (Poland) and OSC (USA).
We are indebted to the communities behind the multiple open-source
software packages on which we depend.
Individual groups or members have received support from
AvH Foundation (Germany);
EPLANET, Marie Sk\l{}odowska-Curie Actions and ERC (European Union);
ANR, Labex P2IO and OCEVU, and R\'{e}gion Auvergne-Rh\^{o}ne-Alpes (France);
Key Research Program of Frontier Sciences of CAS, CAS PIFI, and the Thousand Talents Program (China);
RFBR, RSF and Yandex LLC (Russia);
GVA, XuntaGal and GENCAT (Spain);
the Royal Society
and the Leverhulme Trust (United Kingdom).

%% file: LHCb_Authorship_09-Oct-2019.tex
\centerline
{\large\bf LHCb collaboration}
\begin
{flushleft}
\small
R.~Aaij$^{31}$,
C.~Abell{\'a}n~Beteta$^{49}$,
T.~Ackernley$^{59}$,
B.~Adeva$^{45}$,
M.~Adinolfi$^{53}$,
H.~Afsharnia$^{9}$,
C.A.~Aidala$^{80}$,
S.~Aiola$^{25}$,
Z.~Ajaltouni$^{9}$,
S.~Akar$^{66}$
% ...

{\footnotesize \it

% ...
$ ^{9}$Universit{\'e} Clermont Auvergne, CNRS/IN2P3, LPC, Clermont-Ferrand, France\\
% ...
$ ^{25}$INFN Sezione di Milano, Milano, Italy\\
% ...
$ ^{31}$Nikhef National Institute for Subatomic Physics, Amsterdam, Netherlands\\
% ...
$ ^{45}$Instituto Galego de F{\'\i}sica de Altas Enerx{\'\i}as (IGFAE), Universidade de Santiago de Compostela, Santiago de Compostela, Spain\\
% ...
$ ^{49}$Physik-Institut, Universit{\"a}t Z{\"u}rich, Z{\"u}rich, Switzerland\\
% ...
$ ^{53}$H.H. Wills Physics Laboratory, University of Bristol, Bristol, United Kingdom\\
% ...
$ ^{59}$Oliver Lodge Laboratory, University of Liverpool, Liverpool, United Kingdom\\
% ...
$ ^{66}$Los Alamos National Laboratory (LANL), Los Alamos, United States\\
$ ^{67}$Syracuse University, Syracuse, NY, United States\\
% ...
$ ^{80}$University of Michigan, Ann Arbor, United States, associated to $^{67}$\\
\bigskip
% ...
}
\end{flushleft}